\pdfoutput=1
\documentclass[epj]{svjour}
\RequirePackage{graphicx}
\RequirePackage{mathptmx}      
\RequirePackage{flushend}
\RequirePackage[numbers,sort&compress]{natbib}
\RequirePackage[colorlinks,citecolor=blue,urlcolor=blue,linkcolor=blue]{hyperref}
\usepackage{makecell}
\usepackage{float}
\usepackage{stfloats}
\usepackage{subfigure}
\usepackage{multicol}
\usepackage{multirow}
\usepackage[center]{caption}
\usepackage{comment}
\usepackage{amsmath}
\usepackage{nccmath}
\usepackage{adjustbox}
\usepackage{ulem}
\usepackage{cancel}

\begin{document}
\title{Analysis of $B_c \to \tau\nu_\tau$ at CEPC}
\author{Taifan Zheng\inst{1} \and Ji Xu \inst{2} \and Lu Cao \inst{3} \and Dan Yu \inst{4} \and Wei Wang \inst{2} \and Soeren Prell\inst{5}
  \and Yeuk-Kwan E. Cheung \inst{1} \and Manqi Ruan \inst{4}
 \thanks{\emph{Email:} manqi.ruan@ihep.ac.cn}%
}                     
\offprints{}          
\institute{School of Physics, Nanjing University, Nanjing, China \and
INPAC, SKLPPC, MOE KLPPC, School of Physics and Astronomy, Shanghai Jiao Tong University, Shanghai, China \and
Physikalisches Institut der Rheinischen Friedrich-Wilhelms-Universit\"at Bonn, 53115 Bonn, Germany \and
 Institute of High Energy Physics, Beijing, China \and
Department of Physics and Astronomy, Iowa State University, Ames, IA, USA
}
\date{Received: date / Revised version: date}
%
\abstract{
The precise determination of the $B_c \to \tau\nu_\tau$ branching ratio provides an advantageous opportunity for understanding the electroweak structure of the Standard Model, measuring the CKM matrix element $|V_{cb}|$ and probing new physics models. In this paper, we discuss the potential of measuring the processes of $B_c \to \tau\nu_\tau$ with $\tau$ decaying leptonically at the proposed Circular Electron Positron Collider (CEPC). We conclude that during the $Z$ pole operation, the channel signal can achieve five $\sigma$ significance with $\sim 10^9$ $Z$ decays, and the signal strength accuracies for $B_c \to \tau\nu_\tau$ can reach around 1\% level at the nominal CEPC $Z$ pole statistics of one trillion $Z$ decays assuming the total $B_c \to \tau \nu_\tau$ yield is $3.6 \times 10^6$. Our theoretical analysis indicates the accuracy could provide a strong constraint on the general effective Hamiltonian for the $b \to c\tau\nu$ transition. If the total $B_c$ yield can be determined to $\mathcal{O}(1\%)$ level of accuracy in the future, these results also imply $|V_{cb}|$ could be measured up to $\mathcal{O}(1\%)$ level of accuracy.
%
\PACS{
      {PACS-key}{discribing text of that key}   \and
      {PACS-key}{discribing text of that key}
     } 
} 
\maketitle
\section{Introduction}
\label{sec:1}

Weak decays of heavy mesons not only provide a unique platform to test the electroweak structures of the  Standard Model (SM) but can also shed light on new physics (NP) beyond the SM.
Among different species of heavy mesons,  the $B^+_c$\footnote{The charge conjugate state is implied throughout the paper.} meson, discovered in 1998 by the CDF collaboration \cite{Abe:1998wi,Abe:1998fb}, is of particular interest in this regard. The $B^+_c$ meson has  specific production and decay mechanisms, and accordingly the measurement of its mass, lifetime and decay  branching ratios would help to probe the underlining quark dynamics and determine   SM parameters.

Consisting of two heavy quarks of different types, the $B^+_c$ meson has three decay categories: 1) $b$-quark decay with spectator $c$-quark; 2) $c$-quark decay with spectator $b$-quark; 3) annihilation process (e.g. $B^+_c \to \tau^+\nu_\tau, c\overline{s}$). The purely leptonic decay through the annihilation process is sensitive to the decay constant $f_{B_c}$ and the CKM matrix element $|V_{cb}|$. Such a scheme has been used for the determination of $|V_{cd}|$ and $|V_{cs}|$ in $D^+/D^+_s \to \tau^+\nu_\tau, \mu^+\nu_\mu$ \cite{Zyla:2020}. For $|V_{cb}|$, since the $B^+_c \to \tau^+\nu_\tau$ channel has not been discovered, it is measured using inclusive semileptonic $b \to c$ transitions and the exclusive channel of $\overline{B} \to D^*l\overline{\nu}_l$. However, even if $B^+_c \to \tau^+\nu_\tau$ had been discovered, the decay $\overline{B} \to D^*l\overline{\nu}_l$ would still provide a more precise $|V_{cb}|$ measurement.

In recent years a few discrepancies have been found between the SM predictions and different experimental measurements in the bottom sector, especially in tauonic decay modes of $B$ mesons~\cite{Lees:2012xj,Abdesselam:2019dgh,Aaij:2017uff}. In view of no clear signal in the direct searches of NP to date, the implications in low-energy processes are of great importance. The study of tauonic decay modes of $B$ mesons, mostly $B\to D(^*)\tau\nu$ decays, have indicated some hints for lepton flavor universality violation. While these decay modes are very sensitive to vector/axial-vector type interactions, the (pseudo)scalar type interactions which can be induced  in many popular NP models, e.g., the two-Higgs doublet and leptoquark models are less constrained by them.  Due to the mass hierarchy $m_\tau\ll m_{B_c}$ that results in helicity suppression for $B^+_c \to \tau^+\nu_\tau$ with $V-A$ interactions in the SM, $B_c\to\tau\nu$ has a better sensitivity to the (pseudo)scalar NP interactions \cite{Li:2016vvp,Alonso:2016oyd}.  Therefore, measurement of the branching ratio ${\cal B}(B^+_c \to \tau^+\nu_\tau)$ can be a key in the search for NP. As we will show in Section II, based on  the current knowledge,  NP can affect ${\cal B}(B^+_c \to \tau^+\nu_\tau)$ significantly, which highlights the study of this quantity in the future.

The recently proposed CEPC (Circular Electron Positron Collider) \cite{CEPCStudyGroup:2018ghi} provides an excellent opportunity to measure ${\cal B}(B^+_c \\\to \tau^+\nu_\tau)$. It has a circumference of 100 km and two interaction points. Its primary objective is the precision Higgs study at a center-of-mass-energy ($\sqrt{s}$) of 240 GeV with a nominal production of $10^6$ Higgs. In addition, a dedicated $WW$ threshold scan ($\sqrt{s} = 158-172$ GeV) and the $Z$ factory mode ($\sqrt{s} = 91.2$ GeV) will be operated for electroweak and flavor physics studies. The $Z$ factory will produce up to one trillion $Z$ bosons (Tera-$Z$) in two years, far exceeding LEP's production~\cite{LEP Z production}. Such a huge data sample will enable high precision tests of the SM and allow to study many previously unobservable processes. Furthermore, the clean $e^+e^-$ collision environment and the well-defined initial state compared to hadron colliders are advantages for this analysis at the CEPC. (Super) B factories operating at the $\Upsilon$(4S) center-of-mass-energy are below the energy threshold for $B^+_c$ production. A detailed discussion on the various advantages and prospects on flavor studies at CEPC can be found in \cite{CEPCStudyGroup:2018ghi}. 

In this paper, we discuss the potential of measuring the processes of $B^+_c \to \tau^+\nu_\tau$, $\tau^+ \to e^+\nu_e\overline{\nu}_\tau$ and $\tau^+ \to \mu^+\nu_\mu\overline{\nu}_\tau$ in $Z \to b\overline{b}$ at the CEPC. Important backgrounds are other $Z \to c\overline{c}$ and $Z \to b\overline{b}$, especially the decay of $B^+ \to \tau^+\nu_\tau$ in $Z \to b\overline{b}$ events \footnote{Throughout the paper, all of the $B^+_c/B^+ \to \tau^+\nu_\tau$ events are implied to be $Z \to b\overline{b}$ events containing such decays, unless specified otherwise.}. Both $B^+_c$ and $B^+$ have similar masses and event topologies \cite{Zyla:2020}. The main difference is the lifetime (the $B^+_c$ lifetime is around one third of the $B^+$ lifetime). The L3 experiment at LEP had originally searched for $B^+ \to \tau^+\nu_\tau$ in 1997 with $1.475 \times 10^6$ $Z \to q\overline{q}$ events \cite{Acciarri:1996bv}, and determined ${\cal B}(B^+ \to \tau^+\nu_\tau) < 5.7 \times 10^{-4}$ at 90\% CL. The study did not consider the contribution from $B^+_c \to \tau^+\nu_\tau$. However,  \cite{Mangano:1997md,Akeroyd:2008ac} later argued the $B^+_c \to \tau^+\nu_\tau$ contribution could be comparable to the $B^+ \to \tau^+\nu_\tau$ contribution, and that a similar analysis method could be used to measure $B^+_c \to \tau^+\nu_\tau$. Understanding the $B^+ \to \tau^+\nu_\tau$  background is crucial in this analysis.

We estimate the $B^+_c/B^+ \to \tau^+\nu_\tau$ event yield at the CEPC $Z$ pole as follows. The number of $B^+ \to \tau^+\nu_\tau$ events produced is given by:
\begin{equation}\label{N_B}
\begin{split}
   N(B^\pm \to \tau^\pm\nu_\tau) = & N_Z \times {\cal B}(Z \to b\overline{b}) \times 2 \times f(\overline{b} \to B^+ X) \\
     & \times {\cal B}(B^+ \to \tau^+\nu_\tau) \,,
\end{split}
\end{equation}
where $N_Z $ is the total number of $Z$ bosons produced. The factor two accounts for the quark anti-quark pair. The branching ratios ${\cal B}(Z \to b\overline{b}) = 0.1512 \pm 0.0005 $, $f(\overline{b} \to B^+ X) = 0.408 \pm 0.007$, and ${\cal B}(B^+ \to \tau^+\nu_\tau) = (1.09 \pm 0.24) \times 10^{-4}$ are taken from \cite{Zyla:2020}.
For the $B_c$ production, the theoretical result at next-to-leading order in $\alpha_s$ gives ${\cal B}(Z\to B^\pm_cX) =  7.9\times 10^{-5}$~\cite{Jiang:2015jma}, and our estimate of ${\cal B}(B^+_c \to \tau^+\nu_\tau)$ (see the next section) is $(2.36 \pm 0.19)$\%. These numbers give
\begin{eqnarray}\label{ratio}
 R_{B_c/B} = \frac{N(B^\pm_c \to \tau^\pm\nu_\tau)}{N(B^\pm \to \tau^\pm\nu_\tau)} = 0.28 \pm 0.05,
\end{eqnarray}  
where we use $R_{B_c/B}$ to denote the ratio. Note that the actual uncertainty for $R_{B_c/B}$ is larger since we lack the uncertainty for ${\cal B}(Z\to B^\pm_cX)$. We conduct our analysis with $10^{9}$ simulated $Z$ boson decays including $(1.3 \pm 0.3) \times 10^4$ $B^\pm \to \tau^\pm\nu_\tau$ events. For simplicity and a larger signal dataset for analysis, we assume both $N(B^\pm_c/B^\pm \to \tau\nu_\tau)$ are equal to $1.3 \times 10^4$ and discuss other scenarios at the end, since the results are easily scalable for different values of $R_{B_c/B}$. 

The rest of this paper is organized as follows. Sect. \ref{sec:2} gives  the decay width of $B^+_c \to \tau^+\nu_\tau$ in the SM and estimates the effects in NP scenarios. Sect. \ref{sec:3} introduces the detector, software and the MC-simulated event samples. Sect. \ref{sec:4} presents the analysis method and results. The conclusion is given in Sect. \ref{sec:5}.

\section{$B^+_c \to \tau^+\nu_\tau$ in the SM and in NP models}
\label{sec:2}

\begin{figure}
\begin{center}
\includegraphics[width=0.5\textwidth]{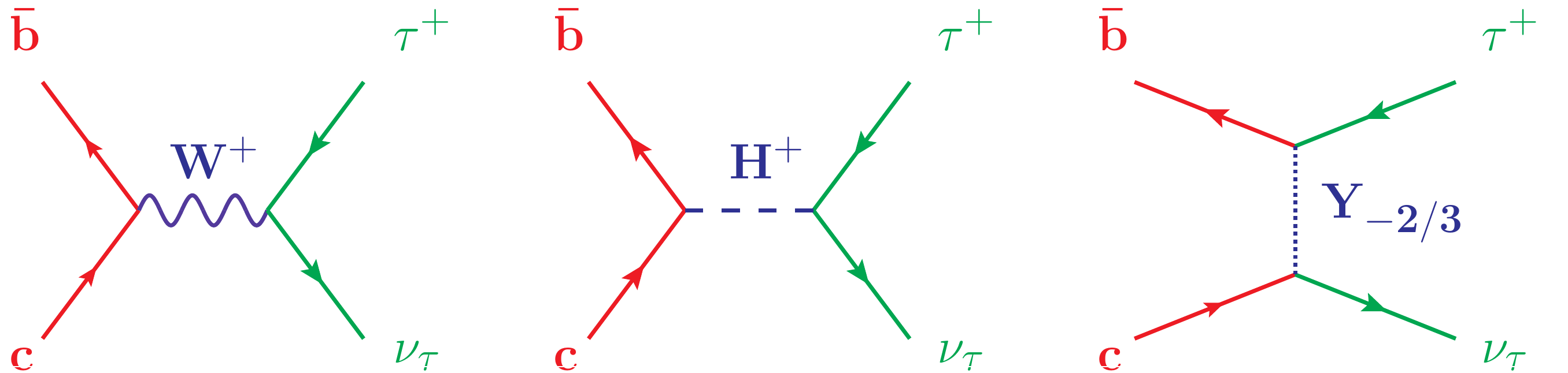}
\caption{Feynman diagrams for tauonic $B_c$ decays in the SM, 2HDM and LQ models.}
\label{Fynmandia}
\end{center}
\end{figure}

In the SM, the decay width of the purely leptonic decay $B^+_c \to l^+\nu_l$ is given by:
\begin{ceqn}
\begin{equation}\label{SM width}
\Gamma_{\textrm{SM}} (B^+_c \to l^+\nu_l) = \frac{G^2_F}{8\pi}|V_{cb}|^2 f^2_{B_c} m_{B_c} m^2_l \left(1 - \frac{m^2_l}{m^2_{B_c}}\right)^2 \,,
\end{equation}
\end{ceqn}
where $G_F$ is the Fermi coupling constant, $V_{cb}$ is the CKM matrix element, $f_{B_c}$ is the decay constant, and $m_{B_c}$, $m_l$ are the masses of the meson and the charged lepton, respectively. Due to helicity suppression, the $\tau$ final state has the largest branching fraction. The measurement of $B^+_c \to \tau^+\nu_\tau$ would help to determine the fundamental parameter $|V_{cb}|$, once the decay constant is known from first-principle calculations, i.e. lattice QCD.  Feynman diagram  for $B^+_c \to \tau^+\nu_\tau$ in the SM is shown in the left panel of Fig. \ref{Fynmandia}.

With the decay constant $f_{B_c}=(0.434\pm0.015)~\textrm{GeV}$~\cite{Colquhoun:2015oha}, $\tau(B_c)= (0.510\pm 0.009)\times 10^{-12}~\textrm{s}$ and $|V_{cb}| = (42.2 \pm 0.8) \times 10^{-3}$ ~\cite{Zyla:2020}, we obtain
\begin{eqnarray}
{\cal B}(B^+_c \to \tau^+\nu_\tau)&=&(2.36\pm0.19)\% \,, \label{eq:BctaunuSM}
\end{eqnarray}
where the errors from the decay constant and lifetime of the $B^+_c$ have been added in quadrature.  The uncertainty in the $B^+_c$ branching fraction is dominated by the decay constant that might be further reduced in a more accurate Lattice QCD calculation in the future. Other theoretical studies on the subject of $B^+_c$ decay can be found in~\cite{Kiselev}.

Since the tau lepton has the largest mass compared to the other two species of leptons, the NP coupling might have a more evident effect in tauonic decays of heavy  mesons. Two popular NP models include the two Higgs doublet model (2HDM) with a charged Higgs boson propagator similar to the $W$ boson propagator,  and the leptoquark (LQ) models that couple leptons with quarks.
The charged Higgs boson in 2HDM can have a significant coupling with the tau,  and thereby its contributions to decay widths could be sizable~\cite{Kalinowski,Hou:1992sy}.

Theoretical studies of NP contributions can be conducted in two distinct ways. One is to confront  the explicit model predictions one by one with available  experimental constraints, while the other is to employ an effective field theory (EFT) approach. Integrating out the massive particles, e.g. charged Higgs particle or the LQ in Fig.~\ref{Fynmandia}, the NP contributions are incorporated  into a few effective operators, with the interaction strengths embedded in  Wilson coefficients. A general effective Hamiltonian for the $b\to c\tau\nu$ transition can be written as
\begin{eqnarray}\label{effectiveh}
  \mathcal{H}_{\mathrm{eff}}&=&\frac{4 G_{F}}{\sqrt{2}} V_{c b}\left[\left(1+C_{V_{1}}\right) {O}_{V_{1}}+C_{V_{2}} {O}_{V_{2}} \right. \nonumber\\
  &&\left.+C_{S_{1}} {O}_{S_{1}}+C_{S_{2}} {O}_{S_{2}}\right]+\mathrm{h.c.} \,,
\end{eqnarray}
where ${O}_i$ are four-fermion operators and $C_i$ are the corresponding Wilson coefficients. The four-fermion operators are defined as
\begin{eqnarray}\label{effectiveo}
  \begin{array}{c}{O}_{V_{1}}=\left(\bar{c}_{L} \gamma^{\mu} b_{L}\right)\left(\bar{\tau}_{L} \gamma_{\mu} \nu_{L}\right), \\
  {O}_{V_{2}}=\left(\bar{c}_{R} \gamma^{\mu} b_{R}\right)\left(\bar{\tau}_{L} \gamma_{\mu} \nu_{L}\right), \\
{O}_{S_{1}}=\left(\bar{c}_{L} b_{R}\right)\left(\bar{\tau}_{R} \nu_{L}\right), \\
{O}_{S_{2}}=\left(\bar{c}_{R} b_{L}\right)\left(\bar{\tau}_{R} \nu_{L}\right).
\end{array}
\end{eqnarray}
where ${O}_{V_{1}}$ is the only operator present in the SM. The 2HDM can contribute to ${O}_{S_{1}}$, while the LQs can have more versatile contributions depending on their spin and chirality in couplings.

Having Eq.(\ref{effectiveh}) and Eq.(\ref{effectiveo}) at hand, one arrives at
\begin{align}
  \frac{\Gamma_{\textrm{eff}}(B^+_c \to \tau^+\nu_\tau)}{\Gamma_{\textrm{SM}}(B^+_c \to \tau^+\nu_\tau)}=\left| 1+C_{V_1}-C_{V_2}+C_{S_1}\frac{m_{B_c}^0}{m_\ell}-C_{S_2}\frac{m_{B_c}^0}{m_\ell} \right|^2 \,,
\end{align}
where $m_{B_c}^0\equiv m_{B_c}^2/(m_b+m_c)$. This expression shows the deviation of decay width of $B^+_c \to \tau^+\nu_\tau$ compared with the SM.

\begin{figure}
\begin{center}
\includegraphics[width=0.45\textwidth]{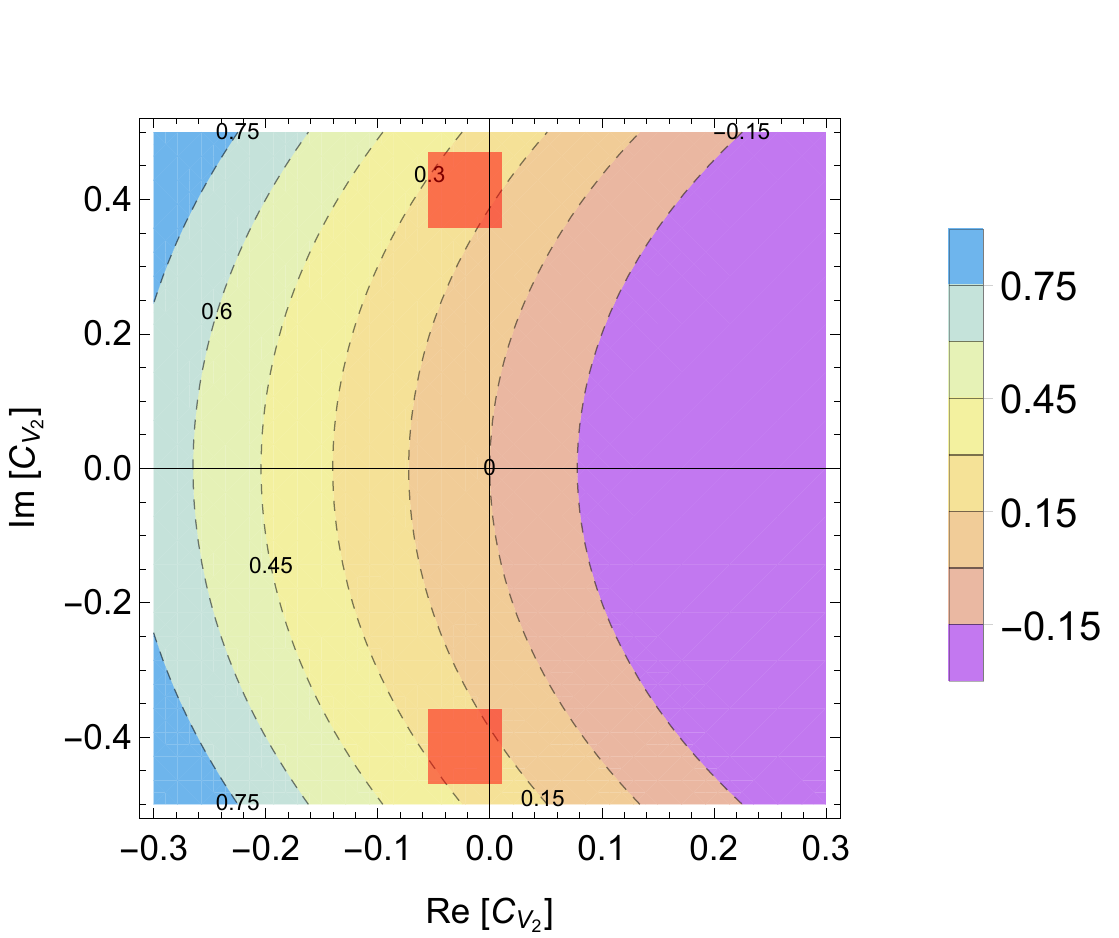}
\caption{Sensitivities of $ (\Gamma_{\textrm{eff}} - \Gamma_{\textrm{SM}})/\Gamma_{\textrm{SM}}(100\%)$ to $C_{V_2}$. The SM lies at the origin with ${\rm Re}[C_{V_2}]={\rm Im}[C_{V_2}]=0$. Labels (in units of $100\%$) on   contours denote the modification of branching ratios (decay widths) with respect to the SM values.  The red shaded area corresponds to the global fitted results of available  data on   $b\to c\tau\nu$ decays, as shown in Eq.~\eqref{eq:CV2}.   These areas deviate from the SM predictions by about a few $\sigma$.  }
\label{BcBr_NP3}
\end{center}
\end{figure}

\begin{figure}
\begin{center}
\includegraphics[width=0.45\textwidth]{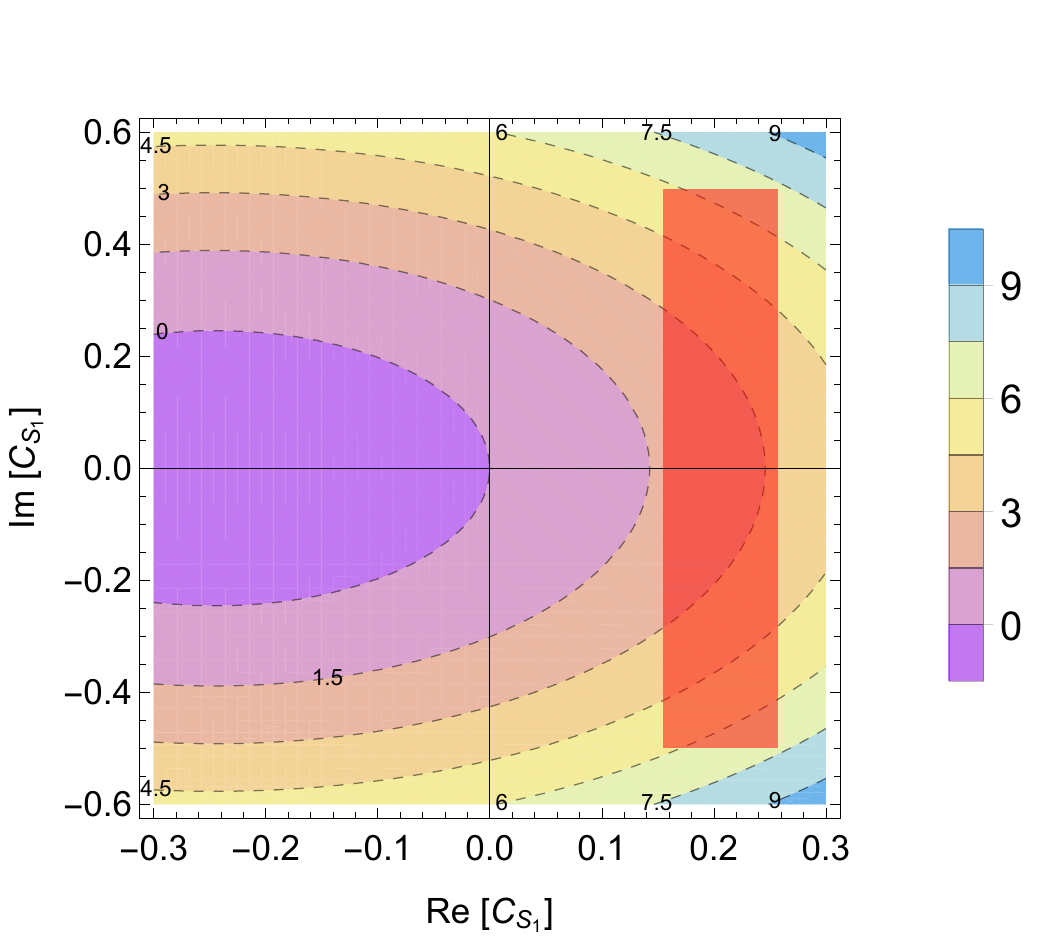}
\caption{Sensitivities of $(\Gamma_{\textrm{eff}} - \Gamma_{\textrm{SM}})/\Gamma_{\textrm{SM}}(100\%)$ to $C_{S_1}$.  The SM lies at the origin with ${\rm Re}[C_{S_1}]={\rm Im}[C_{S_1}]=0$.  Labels (in units of $100\%$) on   contours denote the modification of branching ratios (decay widths) with respect to the SM values.  The red shaded area corresponds to the global fitted results of available  data on   $b\to c\tau\nu$ decays, as shown in Eq.~\eqref{eq:CS1}. }
\label{BcBr_NP1}
\end{center}
\end{figure}

\begin{figure}
\begin{center}
\includegraphics[width=0.45\textwidth]{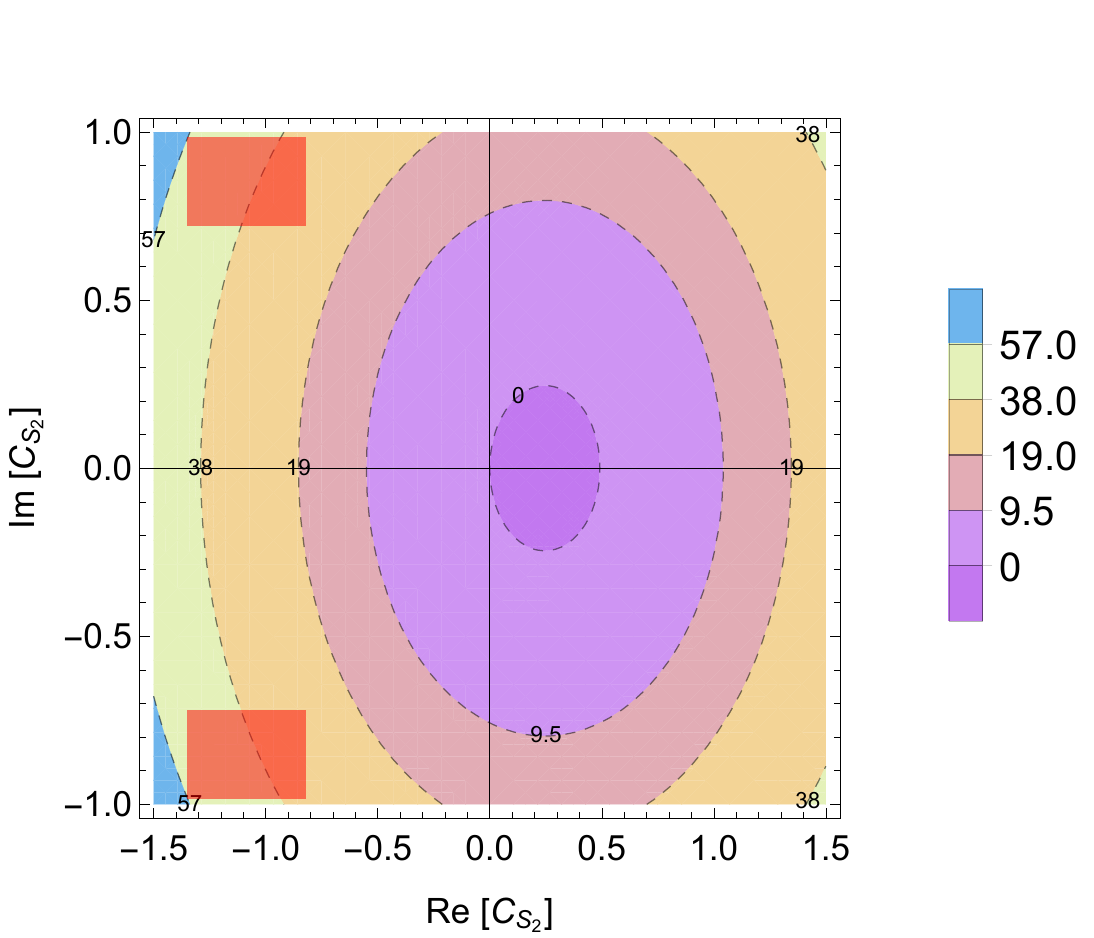}
\caption{Similar to Fig.~\ref{BcBr_NP1} with red shaded area as parameter space of $C_{S_2}$ given in  Eq.~\eqref{eq:CS2}. }
\label{BcBr_NP2}
\end{center}
\end{figure}

Inspired by the experimental measurements of $B\to D(^*)\tau\nu$ and other decays induced by $b\to c \tau\nu$, quite a few theoretical analyses of NP contributions have been made in recent years.  In this work, we will make use of the results for the Wilson coefficients from Refs.\cite{Huang:2018nnq,Cheung:2020sbq}:
\begin{eqnarray}
  && |1+ {\rm Re}[C_{V_1}]|^2 +|{\rm Im} [C_{V_1}]|^2= 1.189\pm0.037 \,,  \label{eq:CV1} \\
  && C_{V_2}= (-0.022\pm 0.033) \pm (0.414\pm 0.056) i \,, \label{eq:CV2} \\
  && C_{S_1}= (0.206\pm 0.051) + (0.000\pm 0.499)i \,,  \label{eq:CS1}\\
  && C_{S_2}= (-1.085\pm 0.264) \pm (0.852\pm 0.132)i, \label{eq:CS2}
\end{eqnarray}
and the masses:
\begin{eqnarray}
  && m_{B_c}= 6.2749 ~\textrm{GeV}\,, \qquad  m_b=4.18 ~\textrm{GeV}\,, \nonumber\\
  && m_c = 1.27 ~\textrm{GeV}\,, \qquad  m_\tau = 1.77686 ~\textrm{GeV}.
 \end{eqnarray}

Eq.~\eqref{eq:CV1} directly implies that the branching fraction of $B^+_c \to \tau^+\nu_\tau$ can be affected by $(18.9\pm3.7)\%$ if only the SM-like $V-A$ operator ${O}_{V_1}$ is included.  If ${O}_{V_2}$ is considered, the contributions to $(\Gamma_{\textrm{eff}} - \Gamma_{\textrm{SM}})/\Gamma_{\textrm{SM}}$ are shown in Fig.~\ref{BcBr_NP3}.  The red shaded area in this figure corresponds to the global fitted results of data on $B$ meson decays induced by $b\to c\tau\nu$, as shown in Eq.~\eqref{eq:CV2}.  In this figure and the following ones, we do not consider the correlation between the real and imaginary part in the Wilson coefficients. Two branches are found due to the ambiguous sign in the imaginary part of $C_{V_2}$.  From this figure, one can infer that the NP contributions range from about $10\%$ to $30\%$.  In these two scenarios,  branching fractions of $B^+_c \to \tau^+\nu_\tau$ are mildly affected due to helicity suppression.

If we switch to ${O}_{S_1}$, the results are shown in Fig.~\ref{BcBr_NP1}, and again the red shaded area corresponds to the global fitted results shown in Eq.~\eqref{eq:CS1}.  Similar results   are shown in Fig.~\ref{BcBr_NP2} for ${O}_{S_2}$.
In these two figures, one can clearly see that $\Gamma(B^+_c \to \tau^+\nu_\tau)$  is dramatically affected by NP contributions. At this stage the errors do not allow a very conclusive result on the existence of NP,  and accordingly measurements of this width at CEPC would help to confirm or rule out these NP scenarios.

Next let's consider the $|V_{cb}|$ measurement in the SM scenario. Its uncertainty can be derived from the relative uncertainty of the signal strength $\sigma (\mu)/\mu$. The signal strength $\mu$ is the ratio between the measured effective cross section and the corresponding SM prediction, and $\sigma(\mu)$ is its uncertainty. Therefore it is straightforward that:

\begin{ceqn}
\begin{eqnarray}\label{eq:Vcb1}
\frac{\sigma (\mu)}{\mu} &=&\frac{\sigma(N(B^\pm_c \to \tau\nu_\tau))}{N(B^\pm_c \to \tau\nu_\tau)} = \frac{\sigma({\cal B}(Z \to B^\pm_cX) {\cal B}(B^+_c \to \tau^+\nu_\tau))}{{\cal B}(Z \to B^\pm_cX) {\cal B}(B^+_c \to \tau^+\nu_\tau)}\nonumber\\
&=&\frac{\sigma({\cal B}(Z \to B^\pm_c) \Gamma_{\textrm{SM}}(B^+_c \to \tau^+\nu_\tau)/\Gamma(B^+_c))}{{\cal B}(Z \to B^\pm_c) \Gamma_{\textrm{SM}}(B^+_c \to \tau^+\nu_\tau)/\Gamma(B^+_c)},
\end{eqnarray}
\end{ceqn}
where $\Gamma(B^+_c)$ is the total width of the $B^+_c$. Substituting Eq.~\eqref{SM width} into the above equation and we have:
\begin{ceqn}
\begin{eqnarray}\label{eq:Vcb2}
 \left( \frac{\sigma(\mu)}{\mu} \right )^2 = \left ( \frac{\sigma({\cal B}(Z \to B^\pm_c X))}{{\cal B}(Z \to B^\pm_c X)} \right )^2 + 4\left ( \frac{\sigma(|V_{cb}|)}{|V_{cb}|} \right )^2 + \nonumber\\
 4\left ( \frac{\sigma(f_{B_c})}{f_{B_c}} \right )^2 + \left ( \frac{\sigma(\Gamma(B^+_c))}{\Gamma(B^+_c)} \right )^2 +  \textrm{Cov.} +\mathcal{O}(10^{-6}),
\end{eqnarray}
\end{ceqn}
where Cov. refers to the covariances between variables. The $\sigma(f_{B_c})/f_{B_c}$ and $\sigma(\Gamma(B^+_c))/\Gamma(B^+_c)$ are both at $\mathcal{O}(1\%)$ level. Sect.~\ref{sec:4} shows that $\sigma(\mu)/\mu$ is also likely at 1\% level at Tera-$Z$. This leaves the error terms to be dominated by the $B^+_c$ production term, which has a much bigger uncertainty, and will determine the uncertainty of $|V_{cb}|$. If the $B^+_c$ production term can be determined to $\mathcal{O}(1\%)$ level in the future and the covariances are also around the same level or less, the $|V_{cb}|$ could be determined to $\mathcal{O}(1\%)$ level as well.

\section{Detector, software and the sample}
\label{sec:3}

The CEPC CDR (conceptual design report) \cite{CEPCStudyGroup:2018ghi} provides a detailed description of the detector setup and the software infrastructure. Both of them are inspired by the International Large Detector (ILD) of the International Linear Collider (ILC) and offer comparable performances. The general flow of software is as follows: 1) create simulated event samples using Pythia \cite{Pythia} and Whizard \cite{Whizard}, 2) the MokkaPlus \cite{Mokka}, a GEANT4 \cite{GEANT4} based simulation tool, simulates the interaction with the detector, 3) the reconstruction framework mimics the electronics' responses and employ Arbor \cite{Arbor}  and LICH \cite{LICH} for physics object creation and lepton identification. Upon completing the standard procedures, two more software are used for further analysis. One is the LCFIPlus \cite{LCFIPlus}, an ILC software which can perform jet clustering and flavor tagging operations to separate different quark flavors in $Z \to q\overline{q}$. The other one is the TMVA \cite{TMVA}, a multi-variable analysis tool for BDT (boosted decision tree) training.

\begin{figure}[h]
  \centering
  \includegraphics[width=0.42\textwidth]{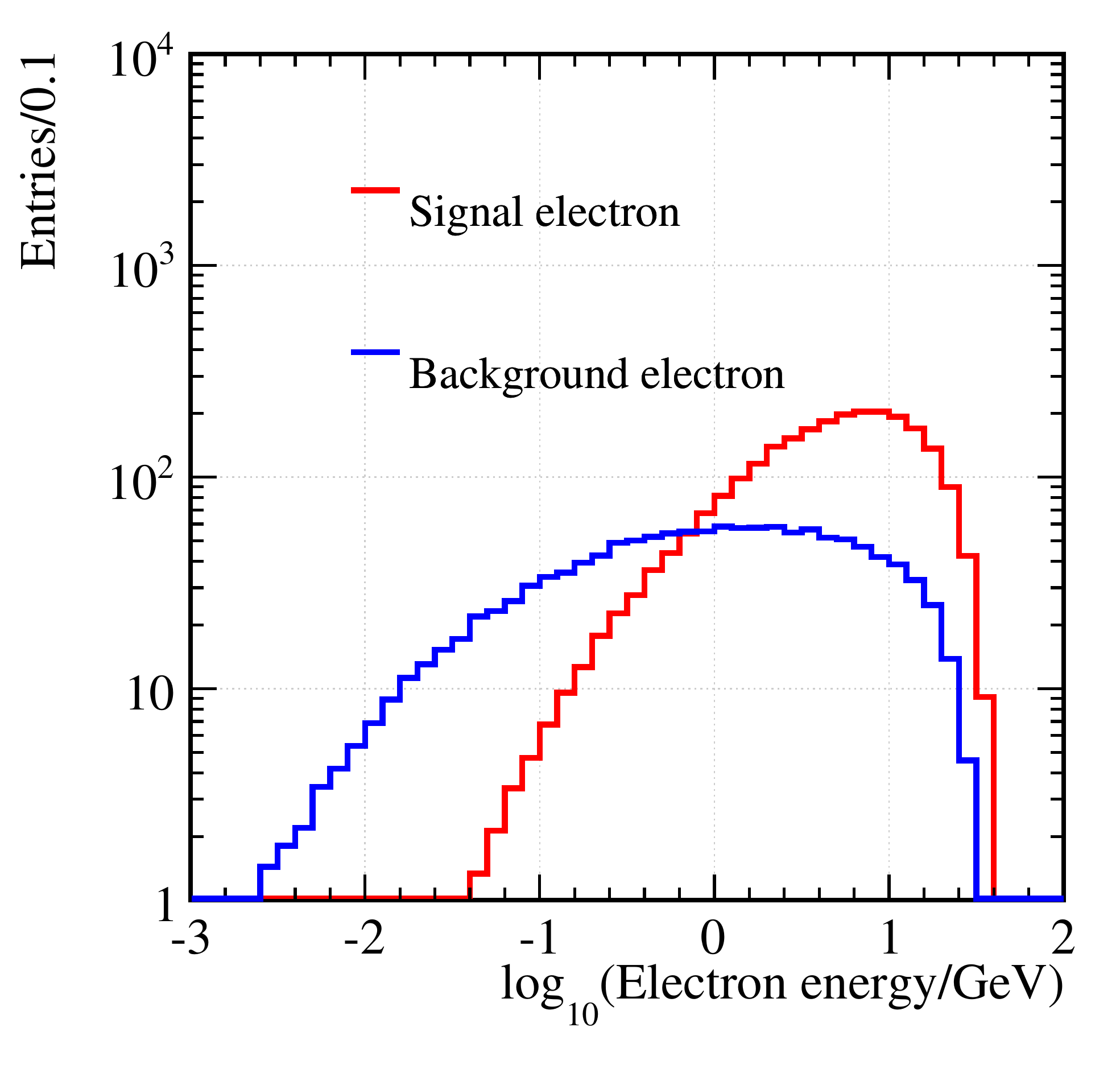}
  \caption{Electron energy distribution in $B_c \to \tau\nu_\tau, \tau \to e\nu\overline{\nu}$.}\label{en}
\end{figure}

The simulated sample consists of $Z \to q\overline{q}, B^+ \to \tau^+\nu_\tau$ and $B^+_c \to \tau^+\nu_\tau$. The latter two are additional $Z \to q\overline{q}$ events that contain the corresponding processes. In order to save time, only a fraction of the $q\overline{q}$ (do not include $B^+_c/B^+ \to \tau^+\nu_\tau$) events that are sufficient for analysis are actually simulated. The data are then scaled to reach the sample size corresponds to $10^9$ $Z$ boson decays. For the $B^+_c/B^+ \to \tau^+\nu_\tau$, we simulated one million events each, and the final numbers and histograms are correspondingly scaled down. All of the scaling factors are shown in Table \ref{cut chain} and Table \ref{muon cut chain}.

Since we are looking for leptonic final states, it is elucidating to demonstrate the lepton identification performance of CEPC. Figure \ref{en} shows the generated energy spectrum of the signal and background electrons from $1.76 \times 10^5$ $B^+_c \to \tau^+\nu_\tau, \tau^+ \to e^+\nu_e\overline{\nu}_\tau$ events (corresponds to one million $B^+_c \to \tau^+\nu_\tau$ events based on the ${\cal B}(\tau^+ \to e^+\nu_e\overline{\nu}_\tau)$. The histograms are scaled down to match $1.3 \times 10^4$ $B^+_c \to \tau^+\nu_\tau$ events.). The signal electrons are the ones from $B^+_c \to \tau^+\nu_\tau, \tau^+ \to e^+\nu_e\overline{\nu}_\tau$. We define the efficiency as the fraction of correctly identified electrons with respect to the total number of electrons.  And the electron mis-identification rate is defined as the rate of hadrons to be identified as electrons \footnote{There is very little cross contamination between electron and muon}. The overall lepton identification efficiency and mis-identification rate at energy above 2 GeV are better than 95\% and 1\%, respectively. For more details, see \cite{LICH}.

\section{Analysis method and results}
\label{sec:4}

\subsection{Analysis method}

\begin{figure}[h]
  \centering
  \includegraphics[width=0.45\textwidth]{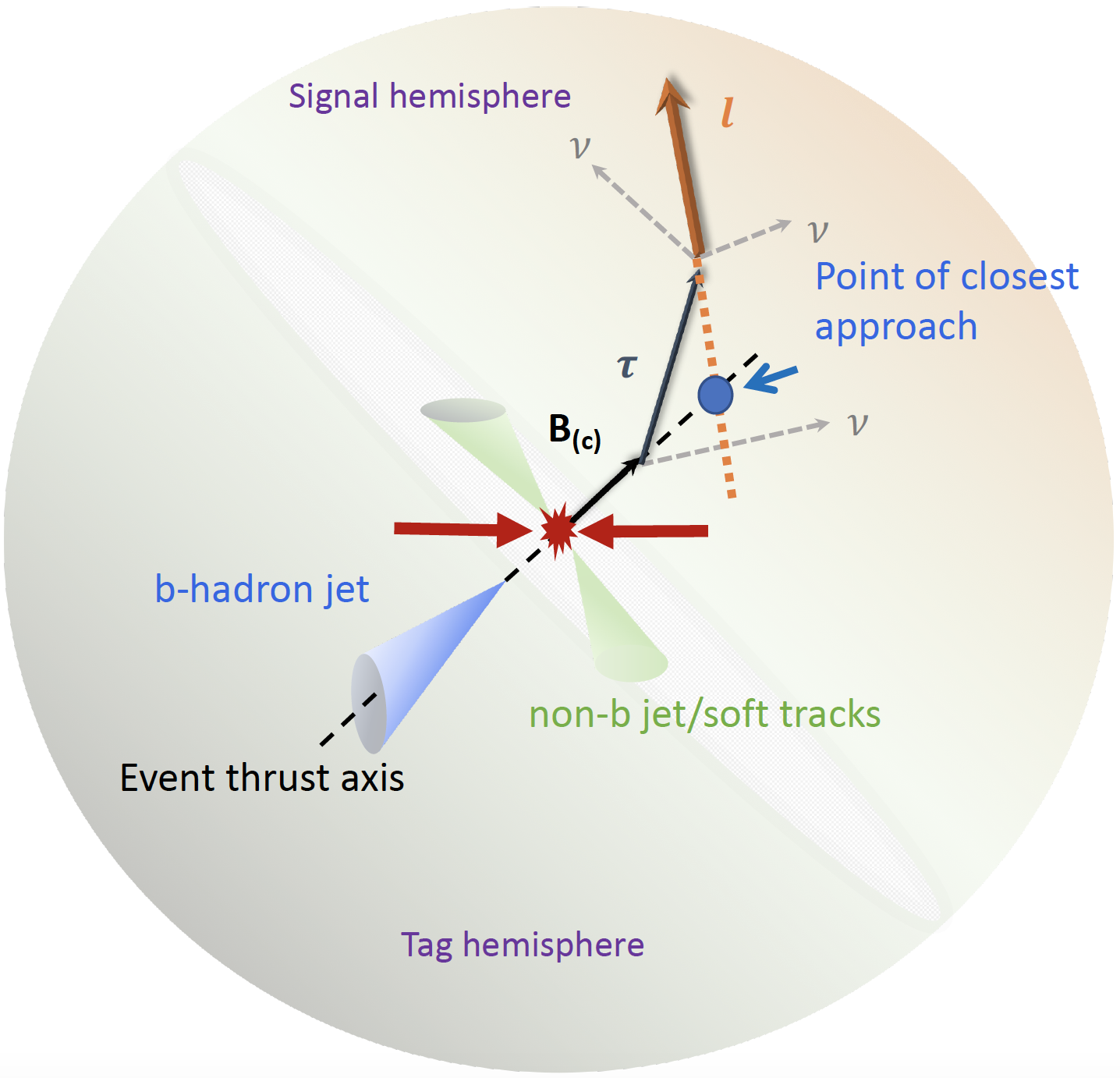}
  \caption{$B_c/B \to \tau\nu, \tau \to e/\mu\nu\overline{\nu}$ in $Z \to b\overline{b}$ event topology. Be reminded that the extension of the lepton track passes close by the thrust axis, but does not need to intersect it.}\label{event topology}
\end{figure}

The characteristic event topology of $B^+_c/B^+ \to \tau^+\nu_\tau, \tau^+ \to e^+/\mu^+\nu\overline{\nu}$ in $Z \to b\overline{b}$ is shown in Fig. \ref{event topology}. The event can be divided into two hemispheres by the plane normal to the thrust. The thrust is the unit vector $\hat{\textit{\textbf{n}}}$ which maximizes
\begin{ceqn}
\begin{equation}
  T = \frac{\Sigma_i|\textit{\textbf{p}}_i \cdot \hat{\textit{\textbf{n}}}|}{\Sigma_i|\textit{\textbf{p}}_i|} \,,
\end{equation}
\end{ceqn}
where $\textit{\textbf{p}}_i$ is the momentum of the $i^\textsuperscript{th}$ final state particle. We let the thrust point towards the hemisphere with less total energy. The axis where the thrust lies is the thrust axis. The hemisphere in which the $B^+_c/B^+ \to \tau^+\nu_\tau, \tau^+ \to e^+/\mu^+\nu\overline{\nu}$ decay occurs is the signal hemisphere and the other one is the tag hemisphere. The main event topology features are: 1) a b-jet in the tag hemisphere, 2) a single energetic $e$ or $\mu$ with relatively large impact parameter along the thrust axis, 3) large energy imbalance between the signal and the tag hemisphere due to missing neutrinos in the signal hemisphere, 4) some soft fragmentation tracks are also present in both hemispheres. Based on the above definitions and features, it is clear that the thrust axis will mostly point towards the signal hemisphere. And the impact parameter is defined as follows: find the point on the thrust axis that is closest to the track, the impact parameter is the signed distance from this point to the interaction point. If the point lies in the signal hemisphere, then the impact parameter is positive, otherwise it is negative. Therefore, the signal lepton's impact parameter characterizes the sum of the decay length of the $B$ meson and the $\tau$. The main difference between $B^+$ and $B^+_c$ events is the impact parameter due to the difference between their lifetimes. The general analysis strategy is:

1. Employ a cut chain which exploits the main features of the event topology to reduce most of the backgrounds from $Z$ decays to light flavor jets.

2. Use a BDT to separate jets with $B^+_c/B^+ \to \tau^+\nu_\tau$, $\tau^+ \to e^+/\mu^+\nu\overline{\nu}$ from other heavy flavor jets. In this case both the $B_c$ and $B$ events are considered as signal.

3. Use another BDT to separate the $B_c$ events from the $B$ and the remaining $b\overline{b}$ events.

Using two BDTs allows us to maximize the separation power of the final state lepton's impact parameter in the second BDT where it will be used as an additional parameter. We begin with the electron final state and later apply the same method to the muon final state as they are highly similar. The first stage cut chain is described in the following:

1. The b-tagging score (ranging from zero to unity) has to be greater than 0.6. This reduces most of non-$b\overline{b}$ $q\overline{q}$ backgrounds.

2. The energy asymmetry, defined as the total energy in the tag hemisphere subtracted by the total energy in the signal hemisphere, has to be larger than 10 GeV. This step significantly reduces all of $q\overline{q}$ events again while preserving most of the $B^+/B^+_c$ events.

3. The signal hemisphere needs to have at least one electron. In case of multiple electrons, the most energetic one is selected for analysis. Most of the signal electrons have sufficient momenta to hit the electromagnetic calorimeter and meet the requirement.

4. The electron is the most energetic particle in the signal hemisphere.

5. The nominal $B$ meson energy is greater than 20 GeV. The quantity is defined as:
\begin{displaymath}
E_B = 91.2\textrm{ GeV} - \textrm{all visible energy except the signal electron}.
\end{displaymath}

Table \ref{cut chain} shows the number of events during the cut chain. We have eliminated most of the light flavor backgrounds. Although their total number is comparable to the signal, considering the corresponding scale factors, they are likely to be eliminated by the following process, hence we ignore the events onwards.

\begin{table*}[ht]
	\centering
	\caption{The cut chain for the electron final state for $10^9$ $Z$ bosons. The numbers in the parentheses are corresponding scale factors. In the final row, the numbers with stars mean the corresponding channels are not used in the second BDT training in order to avoid possible overfitting. Instead, we make a conservative assumption that all of the events passed the first BDT cut survive the second BDT cut.}
    \begin{adjustbox}{center}
	\begin{tabular*}{1\textwidth}{@{\extracolsep{\fill}}cccccccc@{}}\hline
        \multirow{2}{*}{}&\multicolumn{2}{c}{$B^\pm_c \to \tau\nu_\tau$(0.013)}&\multicolumn{2}{c}{$B^\pm \to \tau\nu_\tau$(0.013)}&\multirow{2}{*}{$d\overline{d}(15)$ + $u\overline{u}(12)$ + $s\overline{s}(15)$}&\multirow{2}{*}{$c\overline{c}
        (4.8)$}&\multirow{2}{*}{$b\overline{b}(3.25)$}\\
        &$\tau \to e\nu\overline{\nu}$&excl. $\tau \to e\nu\overline{\nu}$&$\tau \to e\nu\overline{\nu}$&excl. $\tau \to e\nu\overline{\nu}$&&&\\\hline
        All events&	2,303&	10,691&	2,270&	10,633&	419,928,342&	119,954,033&	151,286,603\\
        b-tag $>$ 0.6&	1,611&	7,463&	1,547&	7,151&	2,134,617&		7,344,014&	116,723,067\\
        \makecell[c]{Energy asymmetry\\$>$ 10 GeV}&	1,425&	6,184&		1,389&	5,801&	486,762&	1,609,771&	30,064,030\\
        \makecell[c]{Has electron in\\signal hemisphere}&	1,273&	1,300&	1,243&	1,132&	143,595&	625,670&		15,905,613\\
        \makecell[c]{Electron is the most\\energetic particle}&915&	116&		859&		93&	8,490&	79,190&	4,587,248\\
        $E_B > 20$ GeV&	909&		112&		852&		88&		981&		34,147&	3,203,073\\\hline
        \makecell[c]{$1^\textsuperscript{st}$ BDT score $>$ 0.99}&	390&		12&		259&		4&		---&		48&		910\\\hline
        \makecell[c]{$2^\textsuperscript{nd}$ BDT score $>$ 0.4}&	199&	$12^\star$	&	73&	$4^\star$&		---&	$48^\star$&		33\\\hline
	\end{tabular*}
    \end{adjustbox}
    \label{cut chain}
\end{table*}

After the first stage cut chain, we choose several variables for the BDT to eliminate $b\overline{b}$ and $c\overline{c}$ backgrounds. Some of the variables have been used in the L3 analysis~\cite{Acciarri:1996bv}. They are listed as following:

\begin{itemize}
  \item Nominal $B$ meson energy.
  \item Maximum neutral cluster energy inside a 30 degree cone around the thrust axis in the signal hemisphere.
  \item The largest impact parameter along the thrust axis in the signal hemisphere besides the selected electron. After the cut chain, in most events the signal electron has the largest impact parameter in the signal hemisphere.
  \item Energy asymmetry.
  \item Second largest track momentum in the signal hemisphere.
  \item Electron's energy.
  \item Electron's impact parameter along the thrust axis.
\end{itemize}

We then apply cuts on the outputs of two BDTs as described before. In the first BDT, we use all but the electron's impact parameter along the thrust axis. The parameter will then be added in the second BDT.

\subsection{Results}
\begin{figure}[h]
  \centering
  \includegraphics[width=0.45\textwidth]{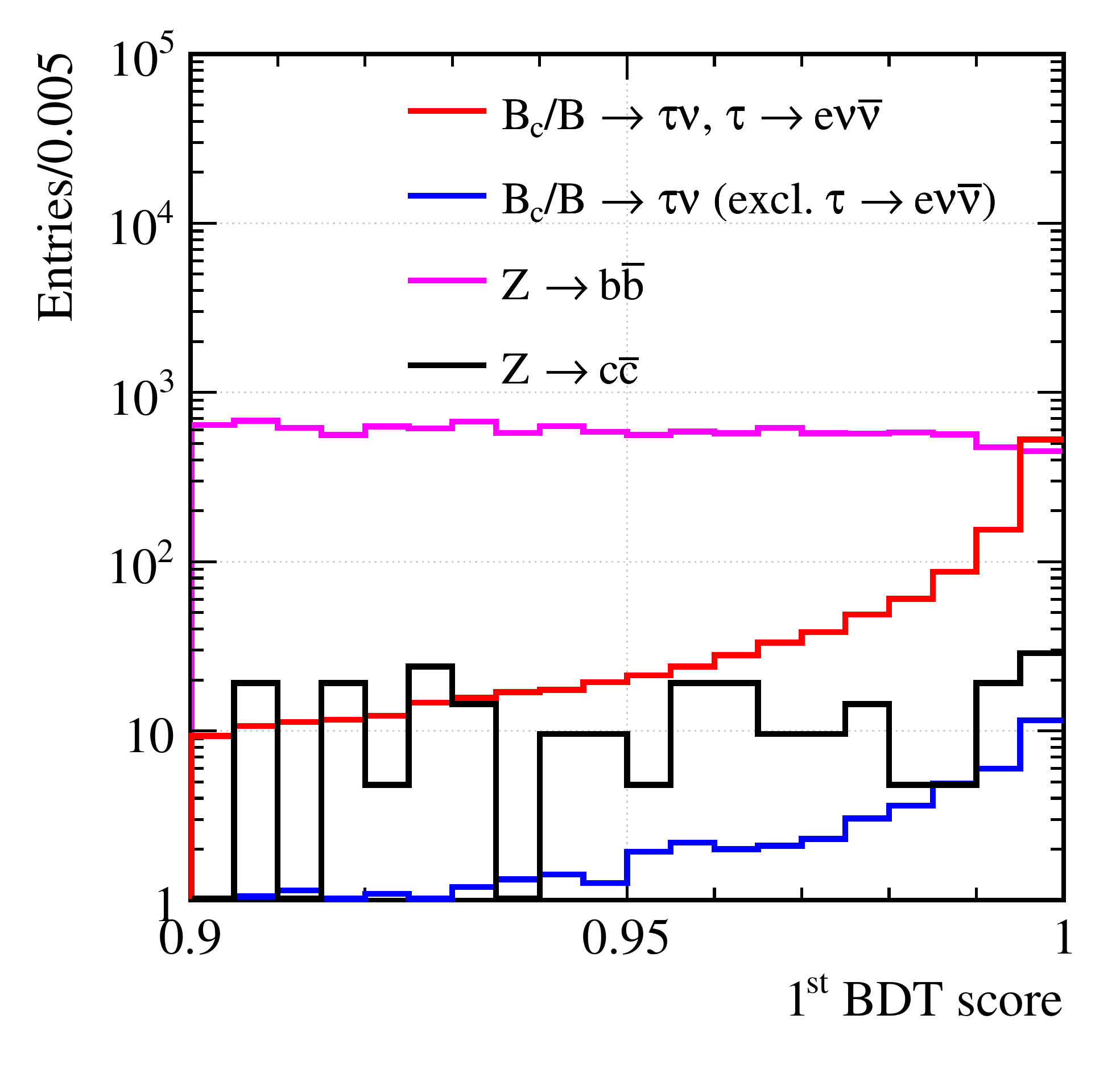}
  \caption{The first BDT score. Here the notation $B_c/B$ means the combination of the two data.}\label{1stBDT}
\end{figure}

\begin{figure}[h]
  \centering
  \includegraphics[width=0.45\textwidth]{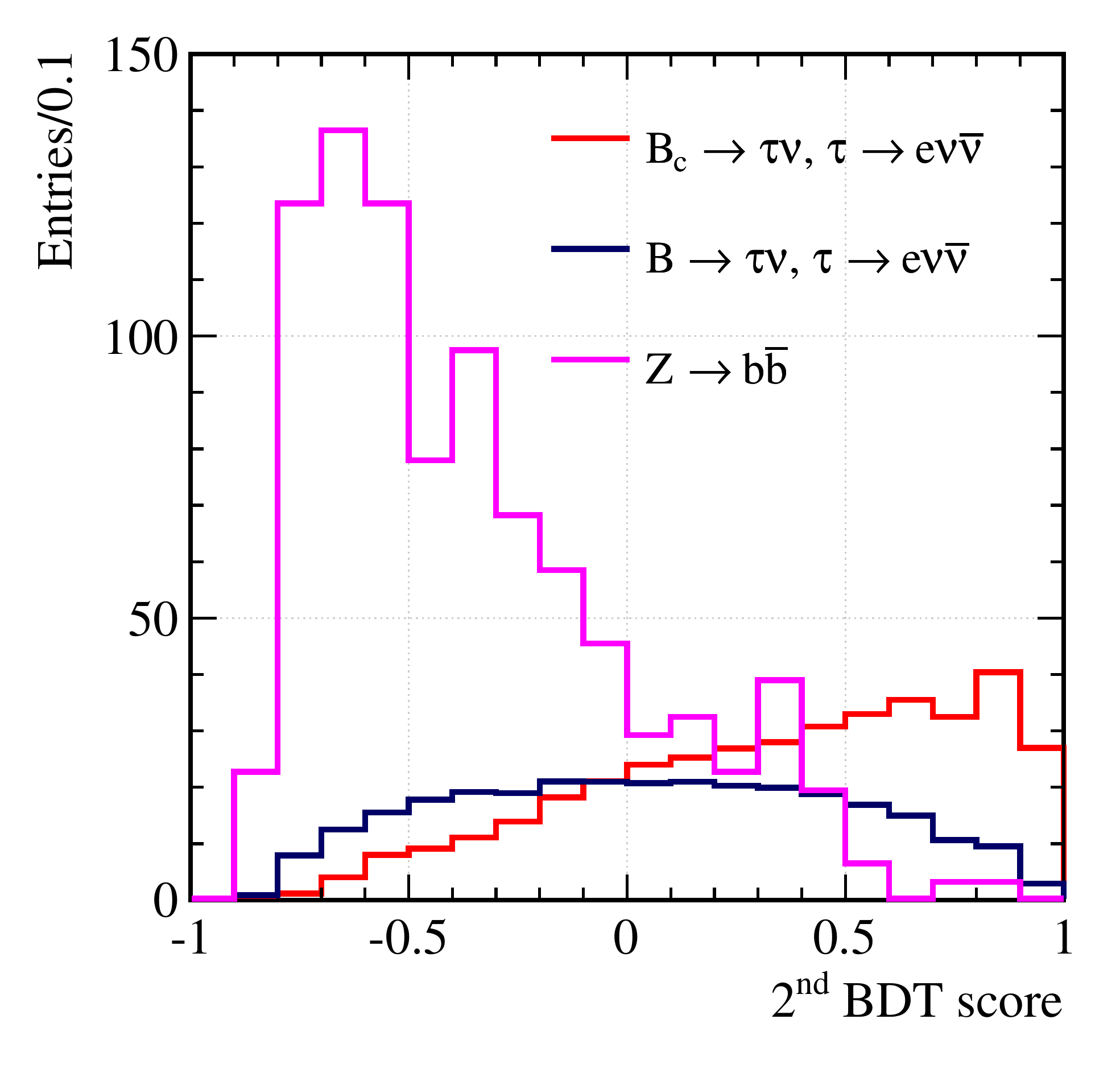}
  \caption{The second BDT score.}\label{2ndBDT}
\end{figure}

The first BDT scores are shown in Fig. \ref{1stBDT}. They range from -1 to 1, of which we showed the rightmost part in the figure. The presence of the signal is apparent at large BDT scores. We apply a cut on the BDT score at 0.99 and only use $B_c/B \to \tau\nu_\tau, \tau \to e\nu\overline{\nu}$ and $Z \to b\overline{b}$ for the second BDT. Ignoring the non-electron $\tau$ decay and $Z \to c\overline{c}$ channels will avoid the possibility of overfitting attributed to these channels, besides the numbers are already small anyway. Then we make a conservative assumption that all of the ignored events survive the second BDT cut, except the light flavor events. The second BDT scores are shown in  Fig. \ref{2ndBDT} and we cut at 0.4. The cut on the BDT scores are chosen to maximize the final signal strength accuracy. Numbers from two BDT results are shown in Table \ref{cut chain}.

\begin{figure}[h]
  \centering
  \includegraphics[width=0.45\textwidth]{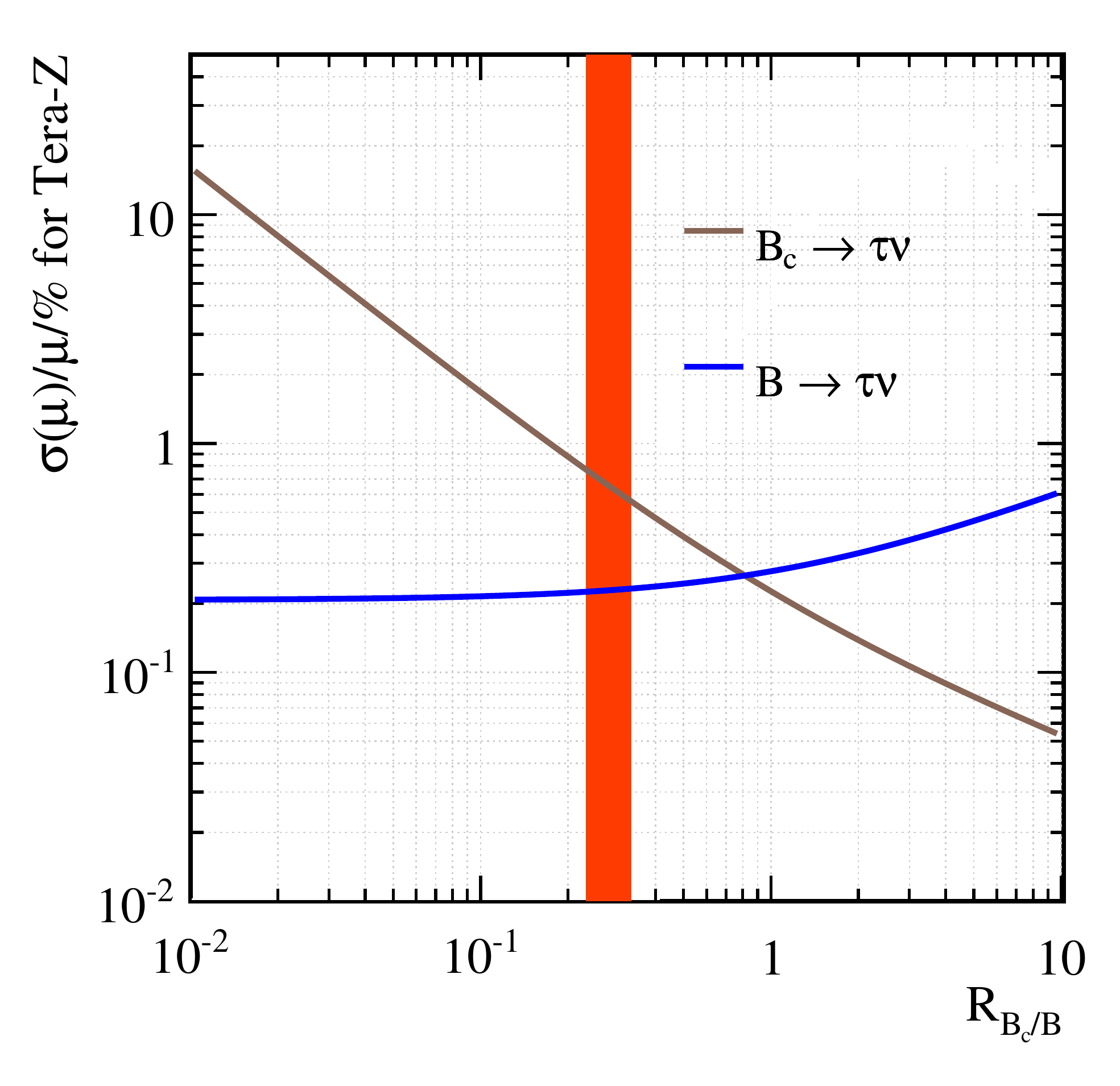}
  \caption{$\sigma(\mu)/\mu$ at Tera-$Z$ versus $R_{B_c/B}$. The estimated range of $R_{B_c/B}$ in Eq. \eqref{ratio} is shown in red band. Be reminded that the actual uncertainty is larger since we lack uncertainty for ${\cal B}(Z \to B^\pm_cX)$.}\label{signal strength}
\end{figure}

Now we can compute the relative accuracy of the signal strength:
\begin{ceqn}
\begin{equation}
  \sigma(\mu)/\mu = \sqrt{N_S + N_B} / N_S \,,
\end{equation}
\end{ceqn}
where $N_S$ and $N_B$ denote the number of signal and background events that pass all selection cuts, respectively.  For the electron final states, we have $\sigma(\mu_e)/\mu_e = 9.7$\%. We can repeat the entire process for the muon final state. Here we will include the non-muon $\tau$ decay channels in the second BDT since the numbers of events are significantly larger. The results are shown in Table \ref{muon cut chain}, and $\sigma(\mu_{\mu})/\mu_{\mu} = 10.6$\%. Combining the two final states, we have $\sigma(\mu)/\mu = 7.2$\%. It is now straightforward to calculate the $\sigma(\mu)/\mu$ for both $B^+_c/B^+ \to \tau^+\nu_\tau$ at Tera-$Z$ at various $R_{B_c/B}$. For the $B \to \tau\nu, \tau \to e/\mu\nu\overline{\nu}$ analysis, all we need to do is repeating the second BDT after switching the signal and background status between it and the $B_c$. Figure \ref{signal strength} shows their relationship with $R_{B_c/B}$. Here, the yield $N(B^\pm \to \tau^+\nu_\tau)$ is fixed at $1.3 \times 10^4$ per one billion $Z$. The projected $\sigma(\mu)/\mu$s at Tera-$Z$ are around $\mathcal{O}(0.1) \sim \mathcal{O}(1)\%$ level for both $B^+_c \to \tau^+\nu_\tau$ and $B^+ \to \tau^+\nu_\tau$. At the $R_{B_c/B}$ value given in Eq. \eqref{ratio}, where the yield $N(B^\pm_c \to \tau^+\nu_\tau)$ is around $3.6 \times 10^3$ per one billion $Z$, we need around $10^9$ $Z$ boson decays to achieve five $\sigma$ significance. In Sect.~\ref{sec:2} we have discussed the $|V_{cb}|$ measurement and with current results we argue that the accuracy could reach up to $\mathcal{O}(1)\%$ level with certain improvements.

\begin{table*}[ht]
	\centering
	\caption{The cut chain for the muon final state for $10^9$ $Z$ bosons. The numbers in the parentheses and the star at the final row have the same meaning as in Table \ref{cut chain}.}
    \begin{adjustbox}{center}
	\begin{tabular*}{1\textwidth}{@{\extracolsep{\fill}}cccccccc@{}}\hline
        \multirow{2}{*}{}&\multicolumn{2}{c}{$B^\pm_c \to \tau\nu_\tau$(0.013)}&\multicolumn{2}{c}{$B^\pm \to \tau\nu_\tau$(0.013)}&\multirow{2}{*}{$d\overline{d}(15)$ +$u\overline{u}(12)$ + $s\overline{s}(15)$}&\multirow{2}{*}{$c\overline{c}
     	(4.8)$}&\multirow{2}{*}{$b\overline{b}(3.25)$}\\
        &$\tau \to \mu\nu\overline{\nu}$&excl. $\tau \to \mu\nu\overline{\nu}$&$\tau \to \mu\nu\overline{\nu}$&excl. $\tau \to \mu\nu\overline{\nu}$&&&\\\hline
        All events&	2,250&	10,745&	2,213&	10,698&	419,928,342&	119,954,033&	151,286,603\\
        b-tag $>$ 0.6&	1,576&	7,499&	1,505&	7,199&	2,134,617&	7,344,014&		116,723,067\\
        \makecell[c]{Energy asymmetry\\$>$ 10 GeV}&	1,387&	6,222&		1,348&		5,848&	486,762&1,609,771&		30,064,030\\
        \makecell[c]{Has Muon in\\signal hemisphere}&	1,175&	2,204&		1,168&		2,233&	244,752&	813,083&		19,569,212\\
        \makecell[c]{Muon is the most\\energetic particle}&882&	222&			838&			171&		9,777&	89,290&		4,943,760\\
        $E_B > 20$ GeV&						877&		216&			832&			166&		1,713&	39,583&		3,516,717\\\hline
        \makecell[c]{$1^\textsuperscript{st}$ BDT score $>$ 0.99}&394&		48&			306&		28&		---&		76&			1,125\\\hline
        \makecell[c]{$2^\textsuperscript{nd}$ BDT score $>$ 0.4}&192&		13&			68&		5&		---&		$76^\star$&			59\\\hline
	\end{tabular*}
    \end{adjustbox}
    \label{muon cut chain}
\end{table*}

\subsection{Phenomenological Impact on New Physics}

\begin{figure}
\begin{center}
\includegraphics[width=0.45\textwidth]{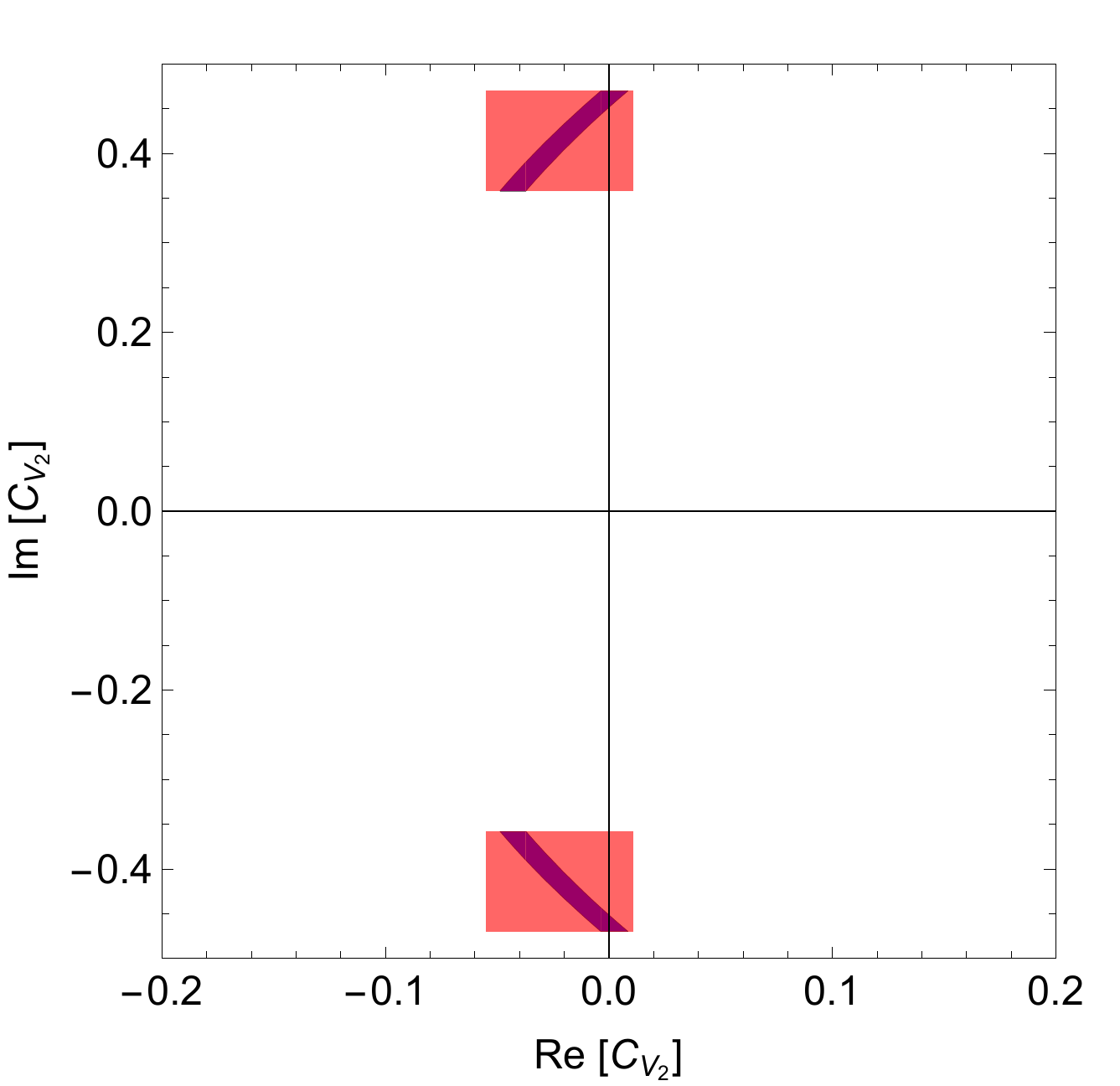}
\caption{Constraints on the real and imaginary parts of $C_{V_2}$.   The red shaded area corresponds to the current constraints using available  data on  $b\to c\tau\nu$ decays.  If the central values in Eq.~\eqref{eq:CV2} remain while the uncertainty in $\Gamma(B^+_c \to \tau^+\nu_\tau)$ is reduced to $1\%$, the allowed region for $C_{V_2}$  shrinks to the dark-blue region.    }
\label{BcBr_NP3-CEPC}
\end{center}
\end{figure}

As we have shown  in Sec.~\ref{sec:2},  based on the current results on NP in   $b\to c\tau\nu$, the   $\Gamma(B^+_c \to \tau^+\nu_\tau)$ tends to deviate from  SM predictions, but the statistical importance is not significant. 
From Fig.~\ref{signal strength}, one can see that at CEPC the $\sigma(\mu)/\mu$ for $B^+_c \to \tau^+\nu_\tau$ can reach about $1\%$ level. This includes the constraint in both the production of $B^+_c$ and the decay into $\tau^+\nu_\tau$. If the production mechanism is well understood, the result on $\sigma(\mu)/\mu$ would also imply that  the uncertainties in $\Gamma(B^+_c \to \tau^+\nu_\tau)$ are reduced to the percent level. On the other side, in the future one can also use the  ${\cal B}(B_c^+ \to J/\psi\pi^+)$ as a calibration mode. In theory the Lattice QCD can calculate the $B_c \to J/\psi$ transition form factors while the perturbative contributions are well under control in perturbation theory. 

One can use such results on  $\Gamma(B_c^+\to \tau^+\nu_\tau)$ to probe NP to a high precision. In Fig.~\ref{BcBr_NP3-CEPC},  we show the constraints on ${\rm Re}[C_{\rm V_2}]$ and ${\rm Im}[C_{\rm V_2}]$.  If the central values in Eq.~\eqref{eq:CV2} remain the same while the uncertainty in $\Gamma(B_c^+\to \tau^+\nu_\tau)$ is reduced to $1\%$, the allowed region for $C_{\rm V_2}$  shrinks as  the dark-blue region, where the  deviation from the SM is greatly enhanced. 

Similar results  can be obtained for NP coefficients $C_{\rm S_1}$ and $C_{\rm S_2}$, but as we have demonstrated in Sec. \ref{sec:2}, both scenarios  will induce dramatic changes to $\Gamma(B_c^+\to \tau^+\nu_\tau)$. These NP effects are so large that they would  already be verified or ruled out before entering into the very precision era of the CEPC. Thus it is less meaningful to present the constraints for these two coefficients.  

\section{Conclusion}
\label{sec:5}

Nowadays hunting for new physics beyond the Standard Model is a primary objective in particle physics.  In this paper, we have first demonstrated that the decay $B^+_c \to \tau^+\nu_\tau$ provides a unique opportunity to probe new physics contributions especially to the (pseudo)scalar interactions that exist in many popular models like the two Higgs doublet model and  the leptoquark models.

We then analyzed the decay $B^+_c \to \tau^+\nu_\tau, \tau^+ \to e^+/\mu^+\nu\overline{\nu}$ at the CEPC $Z$ pole. We took references of the methods used in the L3 analysis \cite{Acciarri:1996bv} on the search of $B^+ \to \tau^+\nu_\tau$, which shares a similar event topology. The backgrounds under consideration are $Z \to q\overline{q}$, $B^+ \to \tau^+\nu_\tau$ as well as other $\tau$ decay channels of $B^+_c \to \tau^+\nu_\tau$. We used a first stage cut chain to suppress most of the light-flavor backgrounds, and subsequently used 2-stage BDT method to perform a fine-tuned multi-variable analysis. The first BDT separates heavy flavor backgrounds and the second BDT separates $B^+ \to \tau^+\nu_\tau$ events. The current detector design and reconstruction algorithms provide excellent signal lepton reconstruction efficiency and purity, and do not pose significant constraints on the analysis. We have demonstrated that under current estimates for $N(B^\pm_c \to \tau^\pm\nu_\tau)$ of around $3.6 \times 10^3$ per one billion $Z$, we need around $\sim10^9$ $Z$ decays to achieve five $\sigma$ significance. The relative accuracy of signal strength could reach around 1\% level at Tera-$Z$. If the total $B^+_c$ yield can be determined to $\mathcal{O}(1\%)$ level accuracy in the future the $|V_{cb}|$ can also be expected to be measured to $\mathcal{O}(1\%)$ level of accuracy. Our theoretical analysis shows the channel has a good potential for NP search and could provide a significant constraint on the NP related to the Wilson coefficient $C_{\rm V_2}$ in Eq. \eqref{effectiveh}. We also showed the projected signal strength accuracy for various signal event numbers for both $B^+_c/B^+ \to \tau^+\nu_\tau$. The results could be improved with a more exhaustive analysis, especially the inclusion of hadronic $\tau$ decays and a larger sample of MC-simulated events.

To summarize, we have demonstrated the CEPC's benchmark capability on the $B^+_c \to \tau^+\nu_\tau$ study. The results show the CEPC could provide a new opportunity to search for the NP such as the 2HDM and LQ models, measure $|V_{cb}|$ and test our understanding of QCD.

\section*{Acknowledgement}
We thank Yiming Li, Haibo Li and Jianchun Wang for useful discussions, and Chengdong Fu and Gang Li for providing some of the samples and tools. We give special thank to Fenfen An for some preliminary studies and useful discussions.
This work is supported by the Beijing Municipal Science \& Technology Commission, project No. Z181100004218003 and Z191100007219010, the Natural Science Foundation of China under grant No. 11735010, 11911530088, 11775110, and 1169\\0034, the Natural Science Foundation of Shanghai under grant No. 15DZ2272100, the DFG Emmy-Noether Grant No. BE 6075/1-1. We also acknowledge the Priority Academic Program Development for Jiangsu Higher Education Institutions (PAPD).

%

\begin{thebibliography}{}
%
%





\bibitem{Abe:1998wi}
F.~Abe \textit{et al.} [CDF],
Phys. Rev. Lett. \textbf{81}, 2432-2437 (1998)
doi:10.1103/PhysRevLett.81.2432
[arXiv:hep-ex/9805034 [hep-ex]].


\bibitem{Abe:1998fb}
F.~Abe \textit{et al.} [CDF],
Phys. Rev. D \textbf{58}, 112004 (1998)
doi:10.1103/PhysRevD.58.112004
[arXiv:hep-ex/9804014 [hep-ex]].




\bibitem{Zyla:2020}
P.A. Zyla et al. (Particle Data Group), to be published in Prog. Theor. Exp. Phys. 2020, 083C01 (2020).




\bibitem{Lees:2012xj}
J.~Lees \textit{et al.} [BaBar],
Phys. Rev. Lett. \textbf{109} (2012), 101802
doi:10.1103/PhysRevLett.109.101802
[arXiv:1205.5442 [hep-ex]].


\bibitem{Abdesselam:2019dgh}
A.~Abdesselam \textit{et al.} [Belle],
[arXiv:1904.08794 [hep-ex]].

\bibitem{Aaij:2017uff}
R.~Aaij \textit{et al.} [LHCb],
Phys. Rev. Lett. \textbf{120} (2018), 171802
doi:10.1103/PhysRevLett.120.171802
[arXiv:1708.08856 [hep-ex]].

\bibitem{Li:2016vvp}
X.~Q.~Li, Y.~D.~Yang and X.~Zhang,
JHEP \textbf{08}, 054 (2016)
doi:10.1007/JHEP08(2016)054
[arXiv:1605.09308 [hep-ph]].


\bibitem{Alonso:2016oyd}
R.~Alonso, B.~Grinstein and J.~M. Camalich,
Phys. Rev. Lett. \textbf{118}, 081802 (2017)
doi:10.1103/PhysRevLett.118.081802
[arXiv:1611.06676 [hep-ph]].





\bibitem{CEPCStudyGroup:2018ghi}
CEPC Study Group,
[arXiv:1811.10545 [hep-ex]].

\bibitem{LEP Z production}
Line Shape Sub-Group of the LEP Electroweak Working Group, DELPHI, LEP, ALEPH, OPAL, L3 Collaboration,
Combination procedure for the precise determination of Z boson parameters from results of the LEP experiments,
[arXiv:hep-ex/0101027[hep-ex]].

\bibitem{Acciarri:1996bv}
M.~Acciarri \textit{et al.} [L3],
Phys. Lett. B \textbf{396}, 327-337 (1997)
doi:10.1016/S0370-2693(97)00138-X

\bibitem{Mangano:1997md}
M.~L.~Mangano and S.~Slabospitsky,
Phys. Lett. B \textbf{410}, 299-303 (1997)
doi:10.1016/S0370-2693(97)00953-2
[arXiv:hep-ph/9707248 [hep-ph]].


\bibitem{Akeroyd:2008ac}
A.~Akeroyd, C.~H.~Chen and S.~Recksiegel,
Phys. Rev. D \textbf{77}, 115018 (2008)
doi:10.1103/PhysRevD.77.115018
[arXiv:0803.3517 [hep-ph]].

\bibitem{Jiang:2015jma} 
  J.~Jiang, L.~B.~Chen and C.~F.~Qiao,
  Phys.\ Rev.\ D {\bf 91}, 034033 (2015)
  doi:10.1103/PhysRevD.91.034033
  [arXiv:1501.00338 [hep-ph]].
  
\bibitem{Colquhoun:2015oha}
B.~Colquhoun \textit{et al.} [HPQCD],
Phys. Rev. D \textbf{91} (2015), 114509
doi:10.1103/PhysRevD.91.114509
[arXiv:1503.05762 [hep-lat]].
  
  \bibitem{Kiselev}
  V.~V.~Kiselev, A.~E.~Kovalsky, A.~K.~ Likhoded,
  Nucl.~Phys.~B585 (2000) 353-382
  doi:10.1016/S0550-3213(00)00386-2
  [arXiv:hep-ph/0002127[hep-ph]].

\bibitem{Kalinowski}
J.~Kalinowski,
Phys.~Lett.~B \textbf{245} (1990) 201-206,
doi:10.1016/0370-2693(90)90134-R.

\bibitem{Hou:1992sy}
  W.~S.~Hou,
  Phys.\ Rev.\ D {\bf 48}, 2342 (1993).
  doi:10.1103/PhysRevD.48.2342

\bibitem{Huang:2018nnq}
Z.~R.~Huang, Y.~Li, C.~D.~Lu, M.~A.~Paracha and C.~Wang,
Phys. Rev. D \textbf{98} (2018), 095018
doi:10.1103/PhysRevD.98.095018
[arXiv:1808.03565 [hep-ph]].

\bibitem{Cheung:2020sbq}
K.~Cheung, Z.~R.~Huang, H.~D.~Li, C.~D.~Lü, Y.~N.~Mao and R.~Y.~Tang,
[arXiv:2002.07272 [hep-ph]].

\bibitem{Pythia}
The Pythia Group, An Introduction to PYTHIA 8.2, Comput. Phys. Commun. \textbf{191} (2015).

\bibitem{Whizard}
W. Kilian, T. Ohl, J. Reuter, WHIZARD: simulating multi-particle processes at LHC and ILC , Eur. Phys. J. C \textbf{71}, 1742 (2011).

\bibitem{Mokka}
C.D. Fu, Full simulation software at CEPC, \url{http://cepcdoc.ihep.ac.cn/DocDB/0001/000167/001}, Accessed 23 Oct 2017.

\bibitem{GEANT4}
S. Agostinelli \textit{et al.}, Geant4-a simulation toolkit. Nucl. Instrum. Methods Phys. Res. Sect. A Accel. Spectrom. Detect. Assoc. Equip. \textbf{506}, 250–303 (2003)

\bibitem{Arbor}
M.Q. Ruan \textit{et al}., Reconstruction of physics objects at the Circular Electron Positron Collider with Arbor, Eur. Phys. J. C \textbf{78}, 426 (2018).

\bibitem{LICH}
D. Yu \textit{et al}., Eur. Phys. J. C \textbf{77} (2017) 591
[arXiv:1701.07542].

\bibitem{LCFIPlus}
T. Suehara, T. Tanabe, LCFIPlus: A framework for jet analysis in linear collider studies, Nuclear Instruments and Methods in Physics Research Section A: Accelerators, Spectrometers, Detectors and Associated Equipment, Feburary, 2016.


\bibitem{TMVA}
A. Hocker \textit{et al}., TMVA-toolkit for multivariate data analysis, physics/0703039, CERN-OPEN-2007-007














\end{thebibliography}
%

\end{document}